\documentclass[seceq]{ptptex}
\usepackage{epsfig,epsf}
\usepackage{bm} 
\newcommand{\beq}{\begin{equation}}
\newcommand{\beql}[1]{\begin{equation}\label{#1}}
\newcommand{\eeq}{\end{equation}}
\newcommand{\bea}{\begin{eqnarray}}
\newcommand{\eea}{\end{eqnarray}}
\newcommand{\eq}[1]{(\ref{#1})}
\newcommand{\fig}[1]{Fig.~\ref{#1}}
\renewcommand{\sec}[1]{Sec.~\ref{#1}}
\newcounter{topiccounter}
\renewcommand{\b}[1]{\mathbf{#1}}
\newcommand{\unit}[1]{\hat {\mathbf{#1}}} 

\newcommand{\as}{\alpha_s}

\title{
Unveiling the nuclear structure at small $x$  using azimuthal correlations in $dA$ collisions%
}

\author{
Kirill \textsc{Tuchin}%
}

\inst{
Department of Physics and Astronomy, Iowa State University, Ames, IA 50011
}

\abst{
We  discuss azimuthal correlations in double inclusive hadron production in high energy $dA$ collisions at RHIC. We argue that the leading logarithmic approximation is inadequate for description of experimental data. Realistic shape of the azimuthal correlation function is obtain only if we keep terms that are finite in the rapidity difference between the produced hadrons. Likewise, transverse momentum dependence of  parton distribution functions in both deuteron and nucleus must be retained. We observe 
 depletion of the back-to-back correlations in central $dAu$ collisions in the forward direction consistent with  the Color Glass Condensate predictions.  
}

\begin{document}

\maketitle

\section{Introduction}\label{sec:intr}

Azimuthal correlations is an important tool to investigate properties of the new form of nuclear matter created in high energy $dA$ and $AA$ collisions at RHIC -- Color Glass Condensate (CGC) -- a coherent quasi-classical state of fast gluons and quarks. In this article we report on a recent study of azimuthal correlations in $dAu$ collisions \cite{Tuchin:2009nf} and demonstrate how the CGC modifies the correlation function. 

In a pioneering paper \cite{Kharzeev:2004bw} it was proposed to study azimuthal correlations of hadrons produced at large rapidity separation $\Delta y\gg 1$. The idea is that such correlations are mediated by the BFKL Pomeron -- a collective state of a large number of gluons, which is a most abandon excitation in the CGC.  Therefore, unlike the hadron production in hard collisions, where there is a strong back-to-back correlation at opening azimuthal  angle $\Delta\phi= \pi$, correlations in the CGC should be significantly reduced. Although only qualitative arguments  were presented in \cite{Kharzeev:2004bw} it seemed that a more quantitative analyses is not too difficult since $\Delta y \gg 1$ corresponds to the well-studied multi-regge-kinematics (MRK) regime and the production amplitude is essentially the real part of the leading order BFKL. We will show however, that the MRK is not sufficient to describe the data at $\Delta y =3$ at RHIC and terms finite in $\Delta y$ must be kept. 

It has been suggested in \cite{Marquet:2007vb} that correlations at small $\Delta y$ in the forward direction can also be used to study CGC. Indeed, forward direction correspond to small $x$ of nucleus where the CGC effects are strongest. They reduce both single and double inclusive hadron production and thus back-to-back correlations are suppressed. 
The problem is that the MRK approximation is not applicable at all in this case.  One therefore has to rely on phenomenological 
models, which offer descriptions that are analytically accurate only in parts of the relevant kinematic region. There are two such approaches: one that is based on the dipole model \cite{Marquet:2007vb,Albacete:2010pg} and another one that is based on the $k_T$-factorization \cite{Tuchin:2009nf}.

In an approrach  based on the `dipole model'  one reduces the relevant scattering amplitudes  to a product of light-cone ``wave-functions" and combinations of CGC field correlators in the configuration space. These correlators satisfy a set of evolution equations with certain initial conditions. The advantage of this approach is that it rests on accurate  theoretical treatment of the gluon saturation region. In this approach the double inclusive gluon \cite{JalilianMarian:2004da}, quark--anti-quark \cite{Tuchin:2004rb,Blaizot:2004wv,Kovchegov:2006qn} and valence quark -- gluon  \cite{Marquet:2007vb} cross sections were calculated. In most cases   in order to obtain analytical results an assumption of  large rapidity gap $\Delta y\gg 1$ must be made. Only in the very forward region dominated by valence quarks of deuteron, i.e.\ at $x_p\sim 1$ finite $\Delta y$ effects have been taken into account \cite{JalilianMarian:2004da,Marquet:2007vb}; transverse momentum dependence of the valence quark distribution is neglected in these papers (resulting in collinear factorization of the deuteron parton distributions). Transverse momentum dependence of valence quark distribution seems to be very important for the shape of the correlation function, though not as important for generation of suppression which is mostly due to large momentum flow from the nucleus. 

 Another approach is `$k_T$-factorization', which assumes that $2\to n$ process and  the two-point correlation functions of CGC fields  can be factored out. In this approximation,  the $2\to 4$ amplitudes were calculated for an arbitrary $\Delta y$ (quasi multi-Regge kinematics, QMRK) in \cite{LRSS,CCH,CE,Fadin:1997hr} for $gg\to ggq\bar q$ and in \cite{Fadin:1996zv,Leonidov:1999nc,Bartels:2006hg} for $gg\to gggg$ processes. 
  Although generally  $k_T$-factorization fails in the gluon saturation region, there are valid reasons to believe that it provides a \emph{reasonable approximation} of the observed quantities. Indeed, it was proved that $k_T$-factorization provides the exact result for the cross section for single inclusive gluon production in the leading logarithmic approximation (LLA) \eq{sincl} \cite{Kovchegov:2001sc} (though there is  a subtlety in the definition of the unintegrated gluon distribution $\varphi$ \cite{Kovchegov:2001sc,Kharzeev:2003wz}). Although $k_T$-factorization fails for the double-inclusive heavy quark production,  the deviation from the exact results is not large at RHIC energies  \cite{Fujii:2005vj}. At  transverse momenta of produced particles much larger than $Q_s$, $k_T$-factorization rapidly converges to the exact results. There are also numerous indications that  $k_T$-factorization is  phenomenologically reliable (see \cite{Tuchin:2009nf} for examples).

\section{Correlations at $|y_T-y_A|\lesssim 1$ }\label{sec2}

First, we would like to consider correlations at small rapidity separations. 
Azimuthal correlation function is defined as 
\beq\label{cf}
C(\Delta \phi)=\frac{1}{N_{\mathrm{trig}}}\frac{dN}{d(\Delta\phi)}\,,
\eeq
where $dN/d(\Delta\phi)$ is the number of pairs produced in the given opening angle $\Delta\phi$ and $N_{\mathrm{trig}}$ is the number of trigger particles. 
The number of pairs is given by
\beq\label{npa}
\frac{dN}{d(\Delta\phi)}= 2\pi \int dk_T k_T \int d y_T\,\int  dk_A k_A \int dy_A
\left( \frac{dN_\mathrm{trig}}{d^2k_Tdy_T}  \frac{dN_\mathrm{ass}}{d^2k_Ady_A}
+ \frac{dN_\mathrm{corr}}{d^2k_Tdy_T\,d^2k_Ady_A}  \right)
\eeq
where $\b k_T$ and $y_T$ are the transverse momentum and rapidity  of the trigger particle and   $\b k_A$ and $y_A$ are the transverse momentum and rapidity  of the associate one. We denote $k_T=\sqrt{\b k_T^2}$ etc.\ throughout this paper.
The first term on the r.h.s.\ of \eq{npa} corresponds to gluon production in two different  sub-collisions (i.e.\ at different impact parameters) and therefore gives a constant contribution to the correlation function, whereas the second term on the r.h.s.\  describes production of two particles in the same sub-collision.  The number of the trigger particles is given by 
\beq\label{strig}
N_\mathrm{trig}= 2\pi \int dk_T k_T\int dy_T\, \frac{dN_\mathrm{trig}}{d^2k_Tdy_T}\,.
\eeq
Expression for the single inclusive  gluon cross section is well-known (see e.g.\ \cite{Kovchegov:2001sc}). The corresponding multiplicity reads
 \beql{sincl}
 \frac{dN}{d^2 k\,dy}\,=\,
\frac{2 \, \as\,}{C_F \, S_\bot}\,\frac{1}{ k^2}\,\int\,
d^2 q_1\varphi_D(x_+,q^2_1)\,\varphi_A(x_-,(\b k- \b q_1)^2)\,.
\eeq
In the center-of-mass frame $x_\pm=\frac{k}{\sqrt{s}}\, \exp\{\pm y\}\,.$
Equation \eq{sincl} is derived in the multi-Regge kinematics (MRK) $x_\pm\ll 1$. 

The correlated part of  double-inclusive parton multiplicity is given by 
\begin{eqnarray}
\frac{dN_\mathrm{corr}}{d^2 k_T\,dy_T\, d^2 k_A\,dy_A}&=&
\frac{N_c \, \as^2}{\pi^2 \, C_F \, S_\bot}\,
\int\,
\frac{d^2 q_1}{q_1^2}\,\int\,\frac{d^2q_2}{q_2^2}\,\delta^2(\b q_1+\b q_2-\b k_T-\b k_A)\nonumber\\
&&\times\,\varphi_D(x_1,q^2_1)\,\varphi_A(x_2,q^2_2)\,\mathcal{A}(\b q_1,\b q_2, \b k_T, \b k_A,y_T-y_A)\,,\label{nlow}
\end{eqnarray}
where $x_{1,2}=(k_{T}e^{\pm y_T} + k_{A}e^{\pm y_A})/\sqrt{s}$.
The amplitude $\mathcal{A}$ was computed in the quasi-multi-Regge-kinematics (QMRK) in \cite{Fadin:1996zv,Fadin:1997hr,Leonidov:1999nc} and recently re-derived in \cite{Bartels:2006hg} (the $gg\to ggq\bar q $ part was calculated before in \cite{LRSS,CCH,CE}). In QMRK one assumes that $x_1,x_2\ll 1$, but $\Delta y $ is finite. 
Explicit expression for $\mathcal{A}$  can be found in  \cite{Leonidov:1999nc}.

For numerical calculations we need a model for the unintegrated gluon distribution function $\varphi$.  In spirit of the KLN model \cite{Kharzeev:2001yq} we write 
\beq\label{kln}
\varphi(x,q^2)=\frac{1}{2\pi^2}\frac{S_\bot C_F}{\as}\big(1-e^{-Q_s^2/q^2}\big) \, (1-x)^4\,.
\eeq 
where the saturation scale of nucleus is $Q_s^2= A^{1/3}Q_{sp}^2$, with $Q_{sp}^2$ the saturation scale of proton fixed by fits of the DIS data. The coupling constant is fixed at $\as=0.3$. 

It has been pointed out in \cite{Leonidov:1999nc} that due to $1\to 2$ gluon splittings the double-inclusive cross section has a collinear  singularity at $\hat s\to 0$, i.e.\ it is proportional to $[(\Delta y)^2  +(\Delta \phi)^2]^{-1}$. Such singularities are usually cured at the higher orders of the perturbation theory. Additional contributions to the small angle correlations arise from various soft processes including resonance decays, hadronization, HBT correlations etc. Because the small angle correlations are beyond the focus of the present paper we simply regulate it by imposing a cutoff on the  minimal possible value of the invariant mass $\hat s$. This is done by redefining the amplitude as $\mathcal{A}\to \mathcal {A} \,\hat s/(\mu^2+\hat s)$. For each kinematic region, parameter $\mu$ is fixed in such a way as to reproduce the value of the correlation function in $pp$ collisions  at zero opening angle $\Delta\phi=0$. 

 $k_T$-factorization is known to give results that are in qualitative agreement with a more accurate approaches, but misses the overall normalization. Therefore, in order to correct the overall normalization of the cross sections we multiply the single inclusive cross section \eq{sincl} by a constant $K_1$ and the double-inclusive one \eq{nlow} by a different constant  $K_2$ \cite{Kovchegov:2002nf,Kovchegov:2002cd}. The correlation function $C$ depends on both $K_1$ and $K_2$. However, the difference $C_\Delta=C(\Delta\phi)-C(\Delta\phi_0)$ depends only on the ratio $K_2/K_1$. We choose $\Delta\phi_0$ in such a way  that $C(\Delta\phi_0)$ is the minimum of the correlation function. This is  analogous to the experimental procedure of removing the pedestal \cite{Adams:2003im}. The overall normalization of  the correlation function $K_2/K_1$ -- which is the only essential free parameter of our model -- is fixed to reproduce the height of the correlation function in $pp$ collisions. 

\begin{figure}[ht]
\begin{tabular}{cc}
      \includegraphics[height=4.5cm]{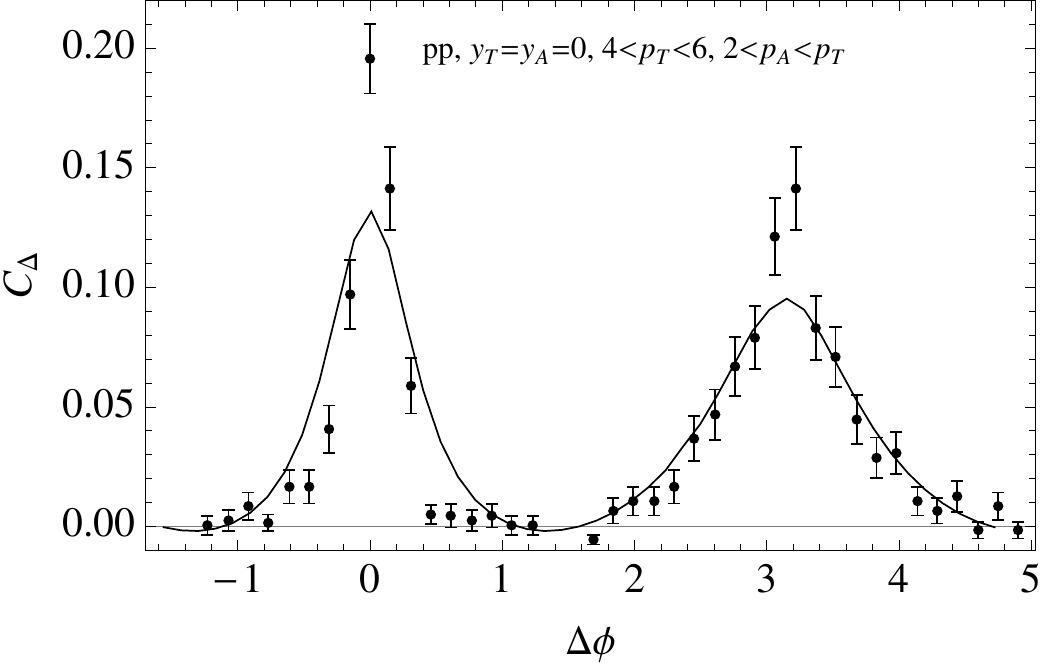} &
      \includegraphics[height=4.5cm]{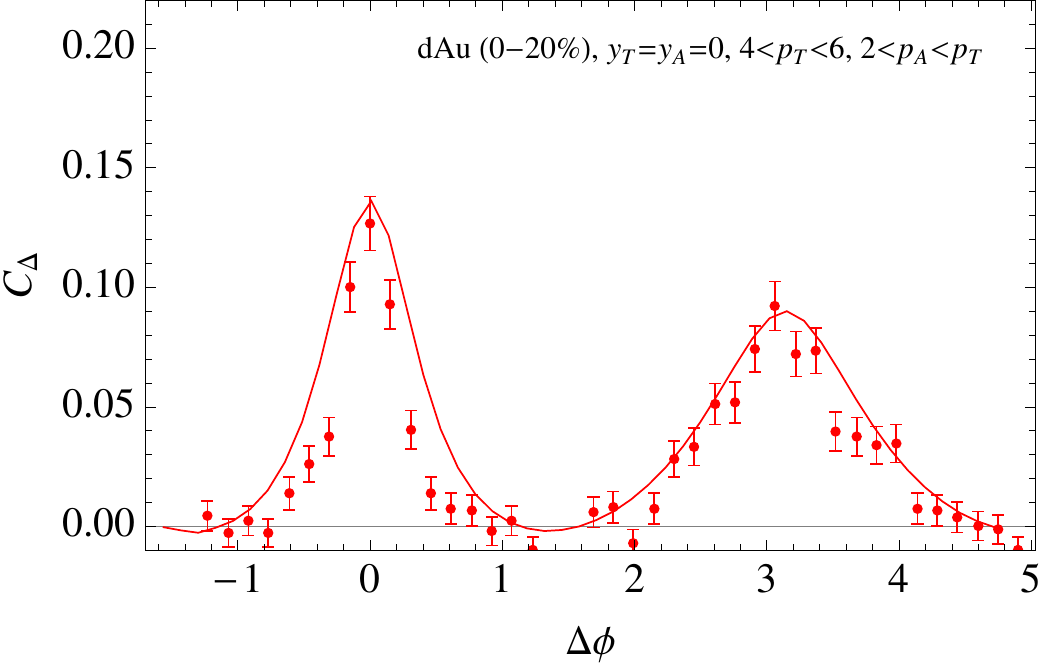} 
  \end{tabular}
  \caption{Correlation function at the central rapidity. Kinematic region is  $4<p_T <6$, $2<p_A <p_T$ (all momenta are in GeV), $y_T=3.1$, $y_A=3$. Left (right) panel: minbias $pp$ ($dAu$) collisions. Data from \cite{Adams:2003im}. }
\label{fig:cc}
\end{figure}

\begin{figure}[ht]
\begin{tabular}{cc}
      \includegraphics[height=4.5cm]{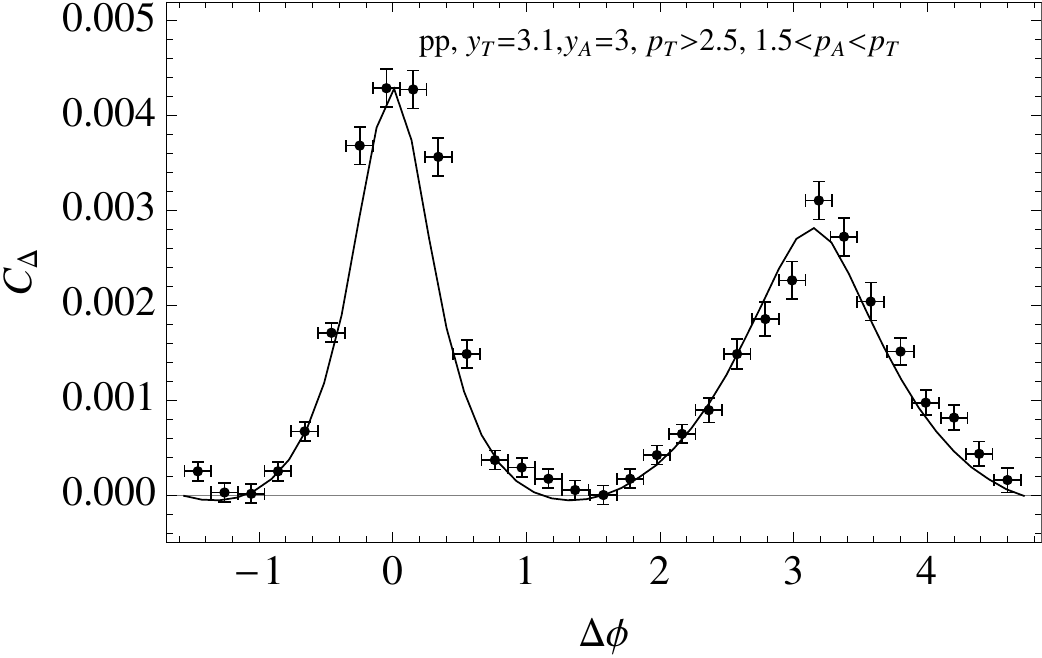} &
      \includegraphics[height=4.5cm]{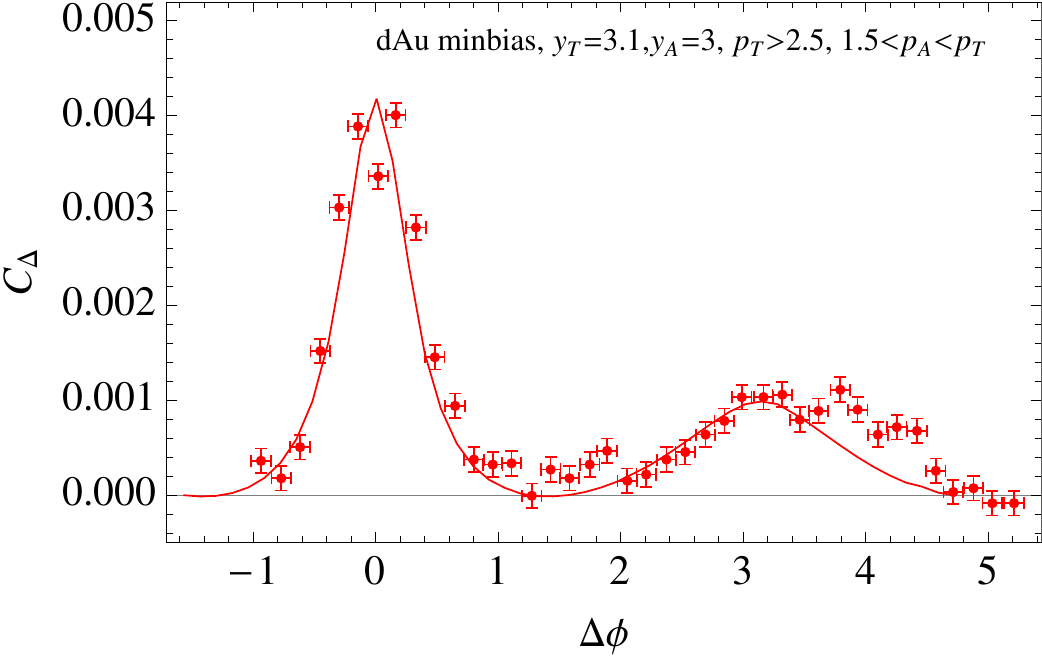} 
  \end{tabular}
  \caption{Correlation function at forward rapidities. Kinematic region is  $p_T>4$, $1.5<p_A <p_T$ (all momenta are in GeV), $y_T=3.1$, $y_A=3$. Left (right) panel:  the minbias $pp$ ($dAu$) collisions. Data from \cite{AGordon}. }
\label{fig:ff1}
\end{figure}

\begin{figure}[ht]
\begin{tabular}{cc}
      \includegraphics[height=4.5cm]{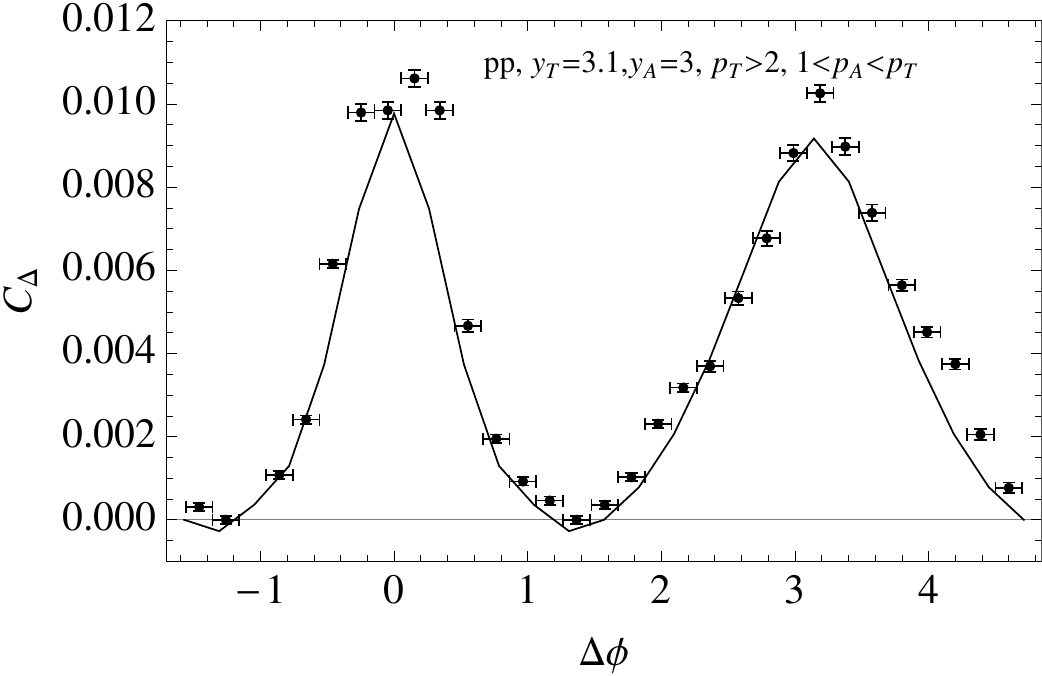} &
      \includegraphics[height=4.5cm]{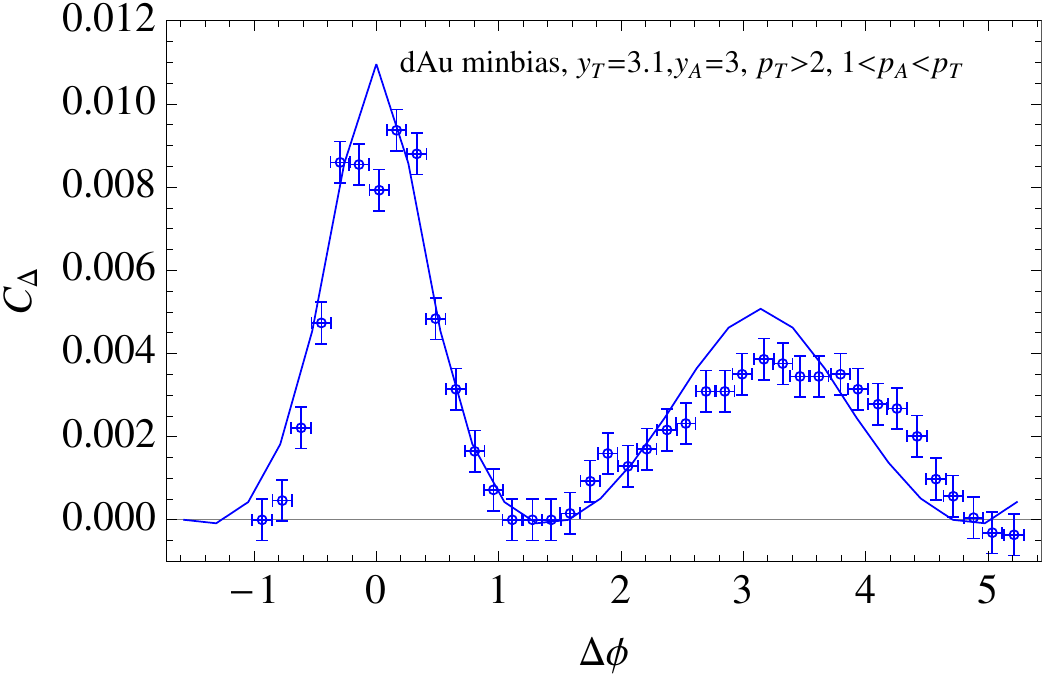} \\
      \includegraphics[height=4.5cm]{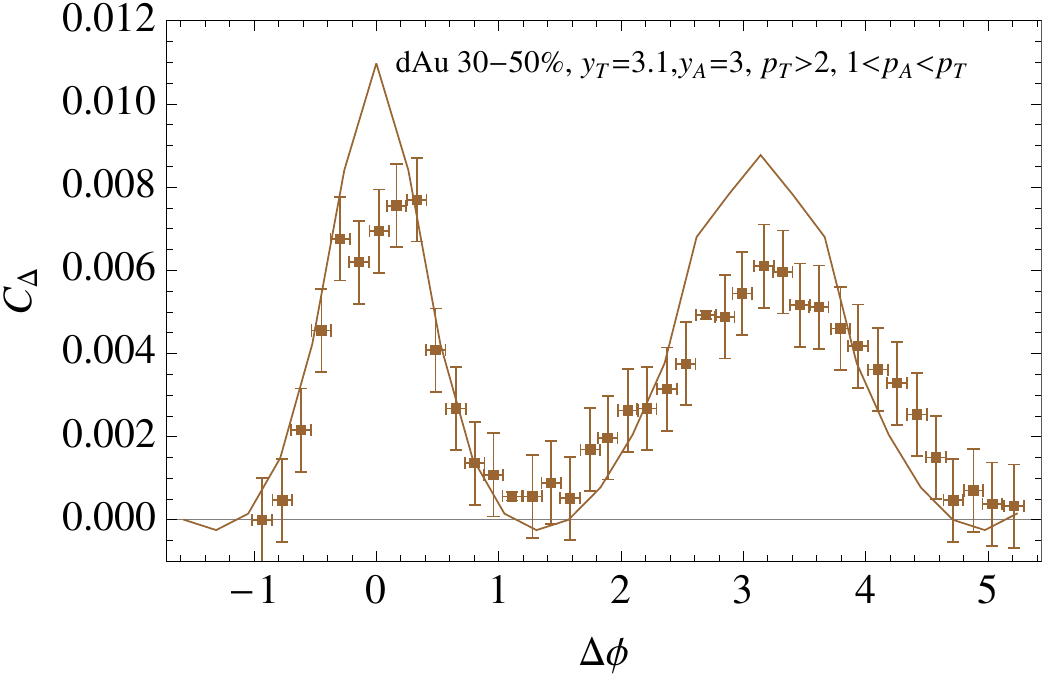}&
      \includegraphics[height=4.5cm]{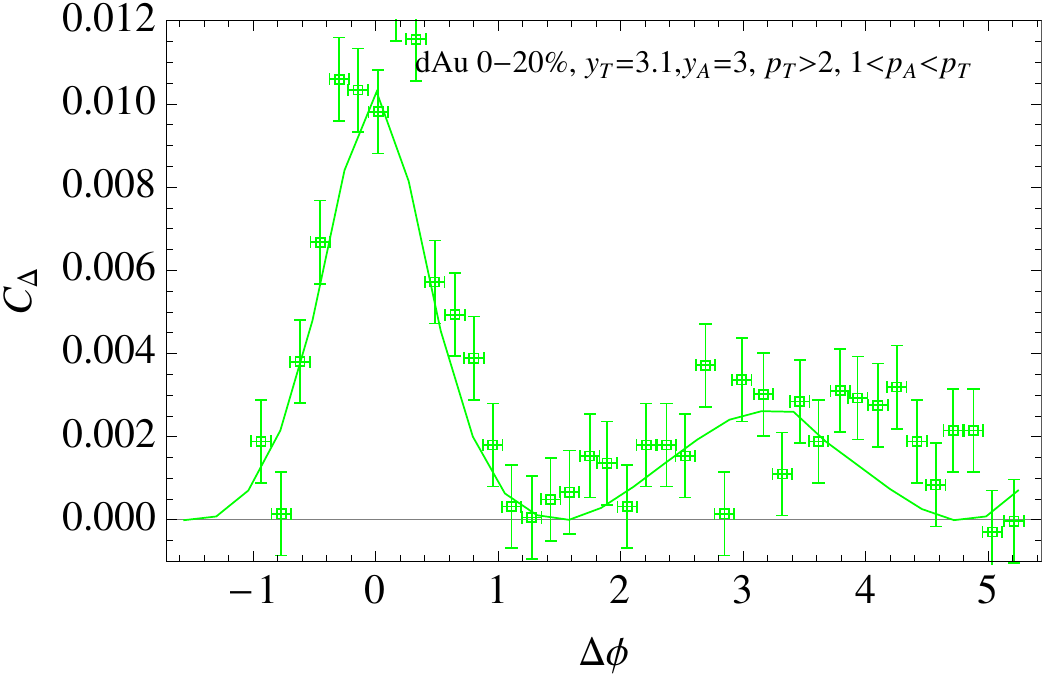}
  \end{tabular}
  \caption{Correlation function at forward rapidities. Kinematic region is  $p_T>2$, $1.5<p_A <p_T$ (all momenta are in GeV), $y_T=3.1$, $y_A=3$. Upper left (right) panel: minbias $pp$  ($dAu$) collisions. Lower left (right) panel: peripheral (central)  $dAu$ collisions. Note: centrality of the theoretical calculation may not coincide with the centrality of the data (the former is not yet known at the time of publication).   Data from \cite{AGordon}.}
\label{fig:ff2}
\end{figure}

The results of the numerical calculations are shown in \fig{fig:cc},\fig{fig:ff1} and \fig{fig:ff2}.  In these figures we observe suppression of the bak-to-back correlation in $dAu$ as compared to the $pp$ ones in agreement with the  experimental data. In \fig{fig:ff2} we also see the depletion of the  back-to-back correlation as a function of centrality. Note, that at the time of publication the precise centrality classes of the \emph{data} shown in the lower row of \fig{fig:ff2} were not known.     

In addition to  $gg\to gggg$ and $gg\to ggq\bar q$ processes that we took into account in this section, production of valence quark of deuteron $gq_v\to gq_v gg$ gives a sizable contribution at forward rapidities due to not very small value of $x$ associated with deuteron ($x\approx 0.2$ for $p_T=2$~GeV at $y=3$). Contribution of this process to azimuthal correlations was analyzed in \cite{Marquet:2007vb} in the framework of the dipole model in MRK. However, the corresponding expression  in $k_T$-factorization in QMRK is presently unknown thus preventing us from taking it into account in our calculation.  In-spite of this we believe that the general structure of the correlation function as well as its centrality dependence is not strongly affected by the valence quark contribution. We plan to address this problem elsewhere. 

\begin{figure}[ht]
\begin{tabular}{cc}
      \includegraphics[height=4.5cm]{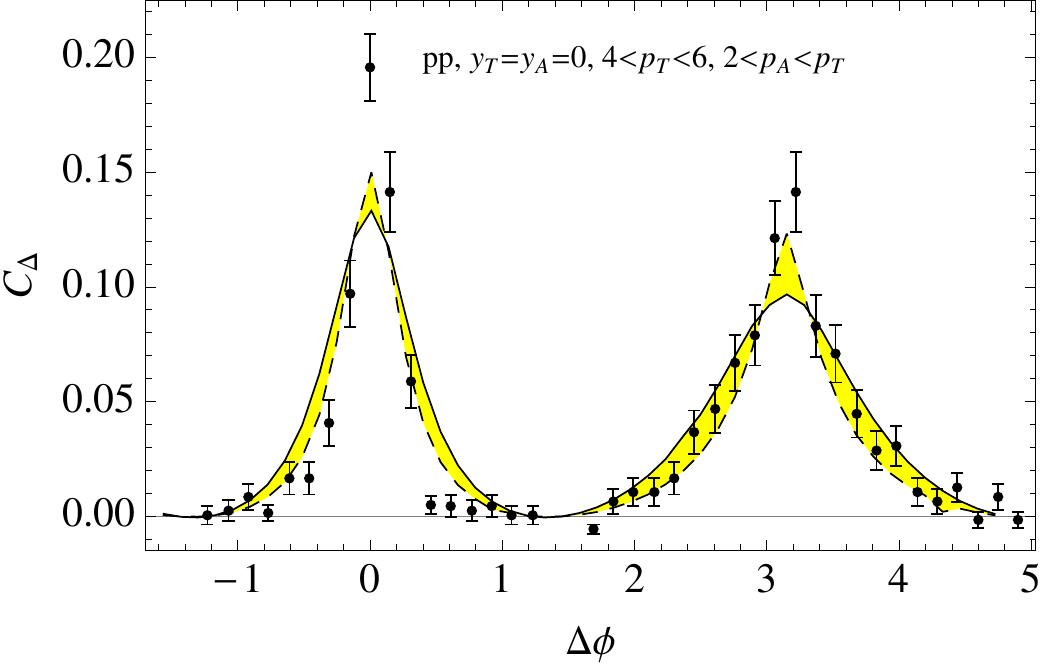} &
      \includegraphics[height=4.5cm]{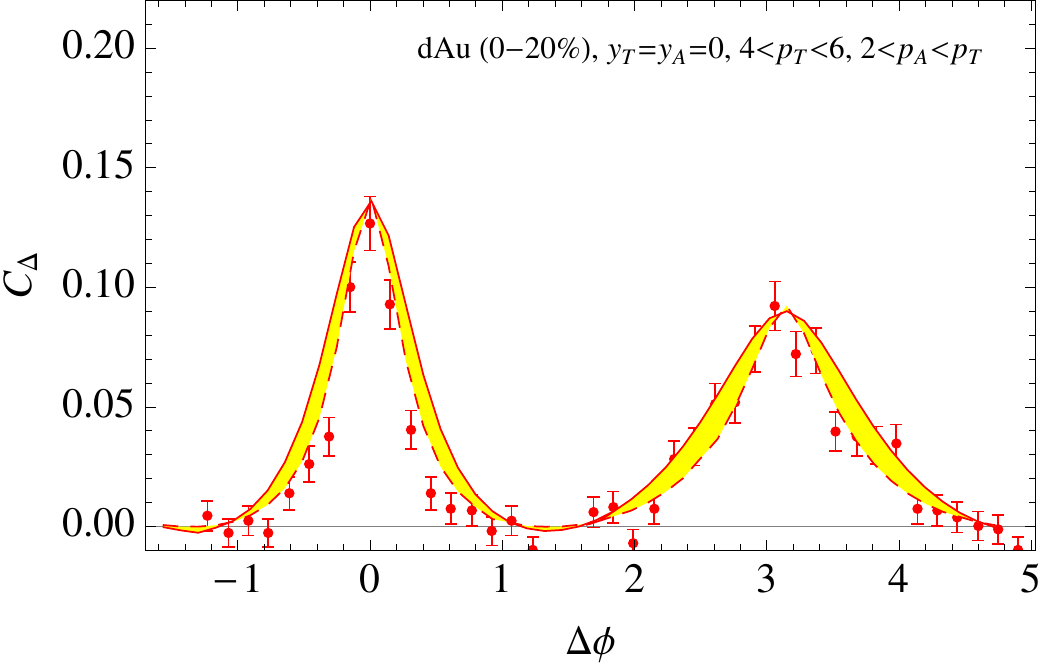} 
  \end{tabular}
  \caption{Effect of fragmentation on the azimuthal correlation function.   Solid lines are the same as in  \fig{fig:cc}. Dashed lines represent a conservative estimate of the fragmentation effect as discussed in the text.
  }
\label{fig:ccfrag}
\end{figure}

So far we neglected the effect of fragmentation. Convoluting the corresponding cross sections with the fragmentation functions taken from \cite{frag} results in minor modifications of  azimuthal distributions. For example, modification of  the correlation function of  \fig{fig:cc} is exhibited in \fig{fig:ccfrag}.
Similar conclusion holds also for the correlation functions at  forward rapidity.

\section{Correlations at $|y_T-y_A|\gg 1$}\label{sec3}

 If the trigger and associated particles are well separated in rapidity so that $|y_T-y_A|\sim 1/\as\gg 1$, we can apply the MRK approximation ($\Delta y \to \infty$) to the double-inclusive cross section. The result then factorizes into a product of two ladder rungs each given by  the real part of the LO BFKL kernel.  The corresponding formula is 
\beql{wincl}
\frac{dN_\mathrm{corr}}{d^2 k_T\,dy_T\, d^2 k_A\,dy_A}=
\frac{N_c \, \as^2}{\pi^2 \, C_F \, S_\bot}\,
\frac{1}{ k_T^2\, k_A^2}\,\int\,
d^2 q_1\,\varphi_D(x_1,q^2_1)\,\varphi_A(x_2,(\b k_T+\b k_A- \b q_1)^2)\,.
\eeq
 The advantage of the MRK approximation  is that it allows taking into account a possible multi-gluon production in the interval between $y_T$ and $y_A$. Unfortunately, in this approximation one also looses many  features of the azimuthal angle dependence that are important for  description of the backward  pick at  intermediate $\Delta y$, see \sec{sec2}. Thus, we are facing a dilemma: 
either to use formulas of \sec{sec2} that give precise dependence on  $\Delta y$ but neglect evolution in the gap, or to take into account the evolution as discussed below but in the MRK limit $\Delta y\to \infty$. At present, there is no approach that would interpolate between these limits at intermediate $\Delta y$ relevant for RHIC. Therefore, in this section we will calculate the correlation function in two limits, compare our results with the data and try to learn which approximation is more phenomenologically important  at $\Delta y=3$.

Eq.~\eq{wincl} does not take into account a  possible gluon emission in the rapidity interval  between  $y_T$ and $y_A$. This is important  when $|y_1-y_2|> 1/\as$ and may be important for the experimentally measured forward-backward rapidity correlations. Evolution in between the  rapidities of the produced particles can be included using the 
 the AGK cutting rules and the known properties of the BFKL equation  as 
 \begin{eqnarray}\label{wev}
\frac{dN_\mathrm{corr}}{d^2 k_T\,dy_T\, d^2 k_A\,dy_A}&=&
\frac{N_c \, \as^2}{\pi^2 \, C_F \, S_\bot}\,
\frac{1}{ k_T^2\, k_A^2}\,\int
d^2 q_1\int
d^2 q_2\,\varphi_D(x_1,q^2_1)\,\varphi_A(x_2,(\b k_T+\b k_A- \b q_1)^2)\nonumber\\
&&\times \,G(|\b q_1-\b k_ T|, |\b q_2-\b k_A|, y_T-y_A)\,,
\end{eqnarray}
where $G$ is the Green's function of the BFKL equation. It can be written as 
\beql{green}
G(q_1,q_2,y)=\sum_{n=0}^\infty 2\cos(n\,\unit q_1\cdot \unit q_2)\, G_n(q_1,q_2,y)\,.
\eeq
where functions $G _n$ are can be found elsewhere (see e.g.\ \cite{Tuchin:2009nf}).

\begin{figure}[ht]
\begin{tabular}{cc}
      \includegraphics[height=4.5cm]{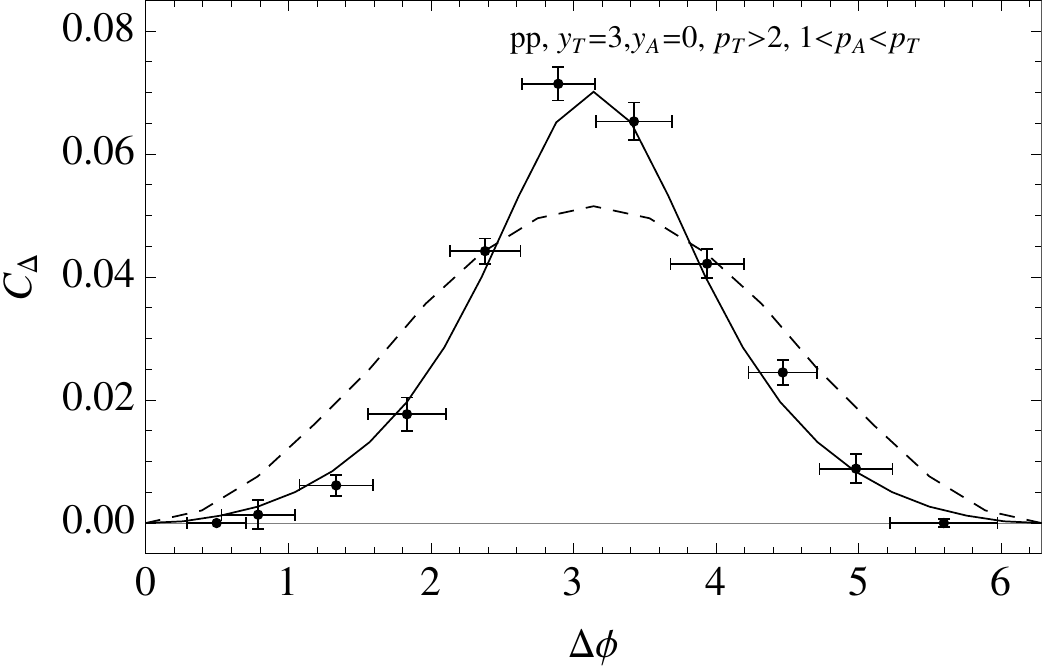} &
      \includegraphics[height=4.5cm]{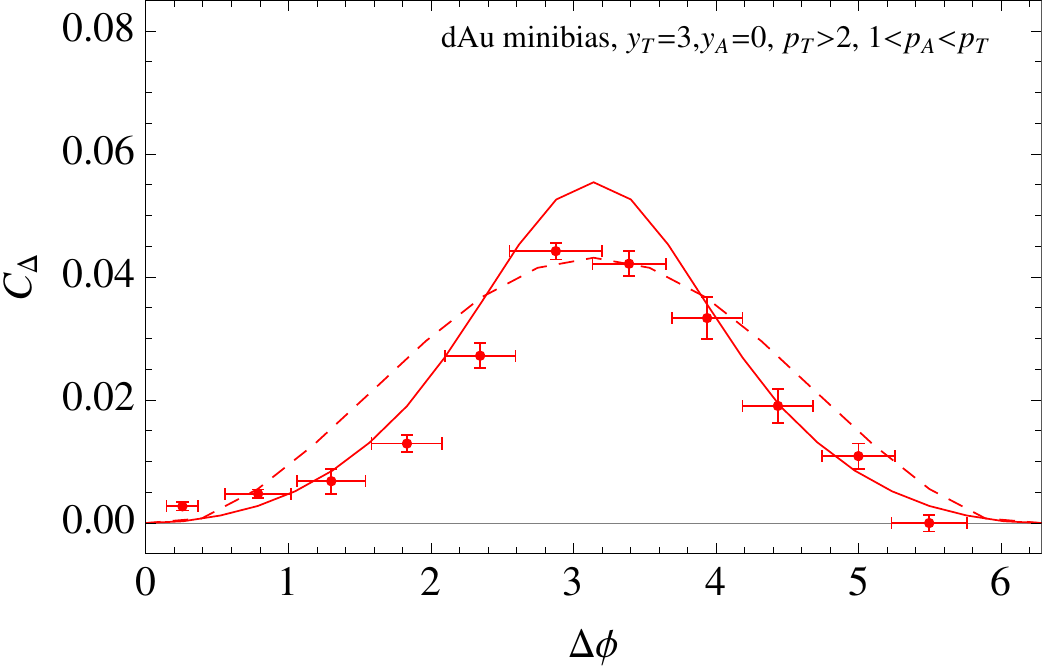} 
  \end{tabular}
  \caption{Forward-backward correlations. Kinematic region is  $p_T >2$, $1<p_A <p_T$ (all momenta are in GeV), $y_T=3$, $y_A=0$. Left (right) panel: minbias $pp$ ($dAu$) collisions. Solid lines: calculations with \eq{nlow} (exact $2\to 4$ amplitude, no evolution between the trigger and associate particles). Dashed line: calculations with \eq{wev} (MRK approximation of $2\to 4$ amplitude, includes evolution between  the trigger and associate particles).  Data  from \cite{Adams:2006uz,Braidot:2009ji} (forward $\pi^0$ and midrapidity $h^\pm$).}
\label{fig:fb}
\end{figure}

It can be easily seen that $G\approx 2[G_0+ G_1\cos(\unit q_1\cdot \unit q_2)]$ is a very good approximation at $\Delta y=3$ \cite{Kharzeev:2004bw}. 
  Numerical calculations are presented in \fig{fig:fb} together with the experimental data from \cite{Adams:2006uz,Braidot:2009ji}. We observe that the shape of the correlation function is better described by \eq{nlow} in agreement with the observation of \cite{Leonidov:1999nc,Kovchegov:2002cd} that  finite $\Delta y$ corrections to the MRK approximation are essential for description of azimuthal correlations. On the other hand, it seems that to explain the magnitude of depletion  one also needs to include the small $x$ evolution effects in the gap $\Delta y=y_T-y_A$. Obviously, a more accurate description requires additional theoretical investigation of the finite $\Delta y$ corrections.    Data  from \cite{Adams:2006uz,Braidot:2009ji} also shows  a significant isospin effect (not displayed here) that probably originates in the valence quark contribution not taken into account in the present work. This isospin effect obscures the CGC contribution and requires a detailed analyses that we plan to do elsewhere. At LHC one can get rid of the isospin effect by considering correlations at large rapidity gaps away from the fragmentation regions.

\section{Conclusions}\label{sec:concl}

We calculated the azimuthal correlation function in $dAu$ collisions using the approach developed by us before \cite{Kovchegov:2002nf,Kovchegov:2002cd}. The results are presented in Figs.~\ref{fig:cc},\ref{fig:ff1},\ref{fig:ff2} and \ref{fig:fb}. We demonstrated that CGC is responsible for depletion of the back-to-back correlations in $dAu$ collisions as compared to those in $pp$ ones
at small rapidity separations -- at midrapidity and forward rapidity -- and at large rapidity separations.  Our results quantitatively confirm earlier arguments of \cite{Kharzeev:2004bw}.

\section*{Acknowledgements}
This work  was supported in part by the U.S. Department of Energy under Grant No.\ DE-FG02-87ER40371. I 
thank RIKEN, BNL, and the U.S. Department of Energy (Contract No.\ DE-AC02-98CH10886) for providing facilities essential
for the completion of this work.

%

\end{document}